\RequirePackage{fix-cm}
\documentclass[twocolumn]{svjour3}          % twocolumn
\smartqed  % flush right qed marks, e.g. at end of proof
\usepackage{graphicx}
\usepackage[dvipsnames]{xcolor}
\usepackage{siunitx}
\usepackage{cite}
\usepackage{xr}
\makeatletter
\newcommand*{\addFileDependency}[1]{% argument=file name and extension
  \typeout{(#1)}
  \@addtofilelist{#1}
  \IfFileExists{#1}{}{\typeout{No file #1.}}
}
\makeatother
\newcommand*{\myexternaldocument}[1]{%
    \externaldocument{#1}%
    \addFileDependency{#1.tex}%
    \addFileDependency{#1.aux}%
}

\myexternaldocument{Suppl}
\journalname{International Journal of Fracture}
\begin{document}

\title{3D characterization of kinematic fields and poroelastic swelling near the tip of a propagating crack in a hydrogel
% in the drained and undrained limits
\thanks{The research was funded by Swiss National Science Foundation Grant no. 200021\_197162.}
}

% \subtitle{Do you have a subtitle?\\ If so, write it here}


\author{Chenzhuo Li         \and
        Danila Zubko        \and
        Damien Delespaul    \and 
        John Martin Kolinski %etc.
}

\authorrunning{C. Li et al.} % if too long for running head

\institute{C. Li \and D. Delespaul \and J. M. Kolinski\at
          School of Engineering, École Polytechnique Fédérale de Lausanne\\
          CH-1015 Lausanne, Switzerland\\
          D. Zubko\at
          School of Computer and Communication Sciences, École Polytechnique Fédérale de Lausanne\\
          CH-1015 Lausanne, Switzerland\\
          \email{john.kolinski@epfl.ch}
}

\date{Received: date / Accepted: date}

\maketitle

\begin{abstract}
In fracture mechanics, polyacrylamide hydrogels have been widely used as a model material for experiments, benefited from its optical transparency, fracture brittleness, and low Rayleigh wave velocity. To describe the brittle fracture in the hydrogels, linear elastic fracture mechanics comes as the first choice. However, in soft materials such as hydrogels, the crack opening can be extremely large, leading to substantial geometric nonlinearity and material nonlinearity at the crack tip. Furthermore, poroelasticity may also modify the local mechanical state within the polymer network. Direct characterization of the kinematic fields and poroelastic effect at the crack tip is lacking.
Here, based on a hybrid method of digital image correlation and particle tracking technique, we retrieved high-resolution 3D particle trajectories near the tip of a slowly propagating crack and measured the near-tip 3D kinematic fields, including the displacement fields, rotation fields, stretch fields, strain fields, and swelling fields. Results confirmed the complex multi-axial stretching near the crack tip and the substantial geometric nonlinearity, particularly on the two wakes of the crack where rotation exceeds $30^{\circ}$. Comparison between the measured and predicted displacement and strain fields, derived from linear elastic fracture mechanics, highlights a disagreement in the direct vicinity of the crack tip, particularly for displacement component $u_x$ and through-thickness strain component $\varepsilon_{zz}$. Significant swelling, due to the poroelastic solvent migration, is also observed, with a strong correlation to the local stretch. Our experimental method, without any assumption of the material properties, can be readily extended to study 3D crack tips in a huge varieties of materials, and our results can shed light on the fundamental fracture mechanics and the development of material models for soft materials undergoing large multi-axial loading and substantial swelling.
% highlight the necessities of the material model development of soft hydrogels considering the 
%  development of , particularly considering 
% Results show that the rotation on the two wakes of the crack becomes significant and can exceed $30^{\circ}$, suggesting a substantial geometric nonlinearity. By evaluating the determinant of the deformation gradient tensor, an increasing swelling is observed approaching the crack tip, indicating a stress-coupled swelling process in this region. 
% Our experimental methods can be extended to study 3D crack tips in a huge varieties of materials, and our results can shed light on the material model development of soft hydrogels.



\end{abstract}

\section{Introduction}\label{sec:intro}

Materials and structures typically fail by fracture, leading to undesired costs and consequences~\cite{anderson2005fracture, freund1998dynamic}. Linear elastic fracture mechanics (LEFM) is the most well-developed theory for brittle fracture, with wide adoption in engineering science~\cite{anderson2005fracture, freund1998dynamic, rice_1968}; it confines the fracture process to a small-scale yielding zone near the crack tip and accurately describes the deformation fields outside this region, with the canonical `$1/\sqrt{r}$' diverging stress field~\cite{anderson2005fracture, freund1998dynamic, williams1952stress, rice_1968}. Within this small region, however, the deformation can be extremely large under the diverging stresses; this large deformation can lead to nonlinear material responses~\cite{livne2008breakdown, bouchbinder2008weakly, Rong2019Fields}, and further induce cohesive loss or poroelastic solvent flux~\cite{li2023pz, deng2023nonlocal, Anand, baumberger2020environmental, yu2018steady, Long_poro, Bouklas_poro, Delayed_gels}. Furthermore, LEFM is developed for predominantly planar cracks that are translationally invariant along $z$, but a real crack can be complex with 3D features~\cite{wei2024complexity, wang2024size, wang2022hidden, tanaka1998discontinuous, baumberger2008magic, kolvin2018topological, ravi1984experimental, fineberg1991instability, sharon1996microbranching, Livne2005, goldman2010, Livne2010Science, sommer, pons_karma}.

For soft materials such as hydrogels, the large deformation results in a exaggerated crack opening~\cite{sun2012highly,kolvin2018supertough,you2020ultra} before fracture, introducing large geometric nonlinearity by local material rotation, despite the experimental merits of hydrogels as a proxy brittle material for fundamental fracture studies\cite{Suo_PAAm_I, Suo_PAAm_II, Suo_PAAm_III, Suo_PAAm_IV, Livne2005, Livne2010Science, goldman2010, wang2023tensile}, and relevance as a bio medical material\cite{yang2008biomaterials, kandow2007bio}. While the incompressible neo-Hookean constitutive law is frequently used as the constitutive model for PAAm hydrogels in many fracture experiments~\cite{Livne2005,  goldman2010, Livne2010Science, wang2024size, wang2022hidden, kolvin2018topological, wei2024complexity, li2023pz}, the stress-induced swelling in polymeric hydrogels is well-known~\cite{baumberger2020environmental, Jin_swelling, Suo_PAAm_V, Suo_diffusion, Chester, chester2, chester3, Anand2, Bouklas_coupled} and confirmed for PAAm hydrogels in uniaxial, biaxial, and indentation experiments~\cite{takigawa1993uniaxial, Urayama2015Biaxial, kalcioglu2012indentation}. Indeed, most of these studies use similar poroelastic models to those used broadly in the earth sciences~\cite{ricecleary, detournay1993fundamentals, detournay2003near, lecampion2018numerical, viesca2021self}. Nevertheless, experiments that probe solvent transport at the tip of a crack, particularly at low-speeds, are lacking; thus, 3D characterization of the kinematic fields at small-scales with high resolution near the crack tip is essential to advance our understanding of poroelastic fracture in hydrogels and the broader class of poroelastic solids.

% In experimental fracture mechanics, polyacrylamide (PAAm) hydrogels are widely used as a model material because of its optical transparency, low Rayleigh wave velocity, facile preparation, and brittleness~\cite{}, alongside its broad application in biological engineering~\cite{}. To extend the measurement into 3D and measure the kinematic fields near the crack tip, the transparency of PAAm hydrogel enables the utilization of optical methods, which offers distinctive merits, including high spatial resolution and simple implementation, over other 3D imaging techniques such as computed tomography~\cite{faccioli2010finger}. .  Using optical method to probe the near-crack-tip kinematic fields in a PAAm hydrogel resolves the 3D imaging challenge and achieves the high-resolution and small-scale requirement.

In this work, we use PAAm hydrogel with an oft-used composition as a model material for poroelastic solids, and embed passive micro particles into the hydrogel as material tracers~\cite{taureg2020dilute}. 3D image stacks of a slowly propagating mode~I crack are obtained using an optical microscope, and the particle trajectories are derived surrounding the crack with subpixel accuracy~\cite{Tyler_F_Estimator, chenouard2014objective, kahler2012resolution}, using a hybrid 3D particle tracking algorithm developed based on open-source software TrackPy~\cite{TrackPy} and Ncorr~\cite{Ncorr}. Based on the particle trajectories, 3D kinematic fields, including the displacement fields, deformation gradient tensor fields, rotation fields, stretch fields, strain fields, and swelling fields, are characterized near the crack tip~\cite{Tyler_F_Estimator}. The in-plane displacement components and strain components are compared with LEFM predictions. Using these kinematic measurements, we analyze the near-crack-tip geometric nonlinearity, triaxial stretch state, and poroelastic solvent migration. Furthermore, the dependency of the solvent flux on crack velocity and the steadiness of the crack speed are discussed. Our experimental methods can be readily extended to 3D complex crack investigations~\cite{wei2024complexity, wang2024size, wang2022hidden, tanaka1998discontinuous, baumberger2008magic, kolvin2018topological, ravi1984experimental, fineberg1991instability, sharon1996microbranching, Livne2005, goldman2010, Livne2010Science, suresh, xu, lazarus, pons_karma, lin, pham1, pham2, dmitry1, brice1, brice2, sommer}, even in other material systems~\cite{Dickey1, Long, Esther, Jin_elastomer, Zehnder_PDMS}.

\section{Method}\label{sec:method}
Experiments are performed on $\SI{450}{\micro m}$-thick polyacrylamide hydrogel samples. The hydrogels are prepared with a precursor of 13.8~wt\% acrylamide monomer and 2.7~wt\% bis-acrylamide cross-linker. The stock solution is first degassed in a vacuum chamber for 10 minutes and mixed with polystyrene particles ($\SI{1.1}{\micro m}$ in diameter) at a concentration of 0.005~wt\%. The mixed solution is sonicated for 5 minutes to disperse the aggregated particles and ensure their uniform distribution. To initiate and accelerate the free-radical polymerization, 0.2\% ammonium persulfate (APS) and 0.02\% tetramethylethylenediamine (TEMED) are added to the hydrogel solution. After mixing for 30 seconds, the solution is poured onto a glass plate and covered with a second glass plate separated by $\SI{380}{\micro m}$-thick spacers, and polymerization reaction proceeds for at least 4 hours.
The resulting polymerized hydrogel is cut into samples of uniform size ($\SI{3}{\cm}$ by $\SI{1}{\cm}$) and soaked in water for 24 hours to reach an equilibrium state. Note that this preparation process, apart from the embedding of the particles, follows a standard protocol used in dynamic fracture experiments~\cite{Livne2005, goldman2010, Livne2010Science}.

\begin{figure*}[!htb]
\centering
\includegraphics[width=0.95\textwidth]{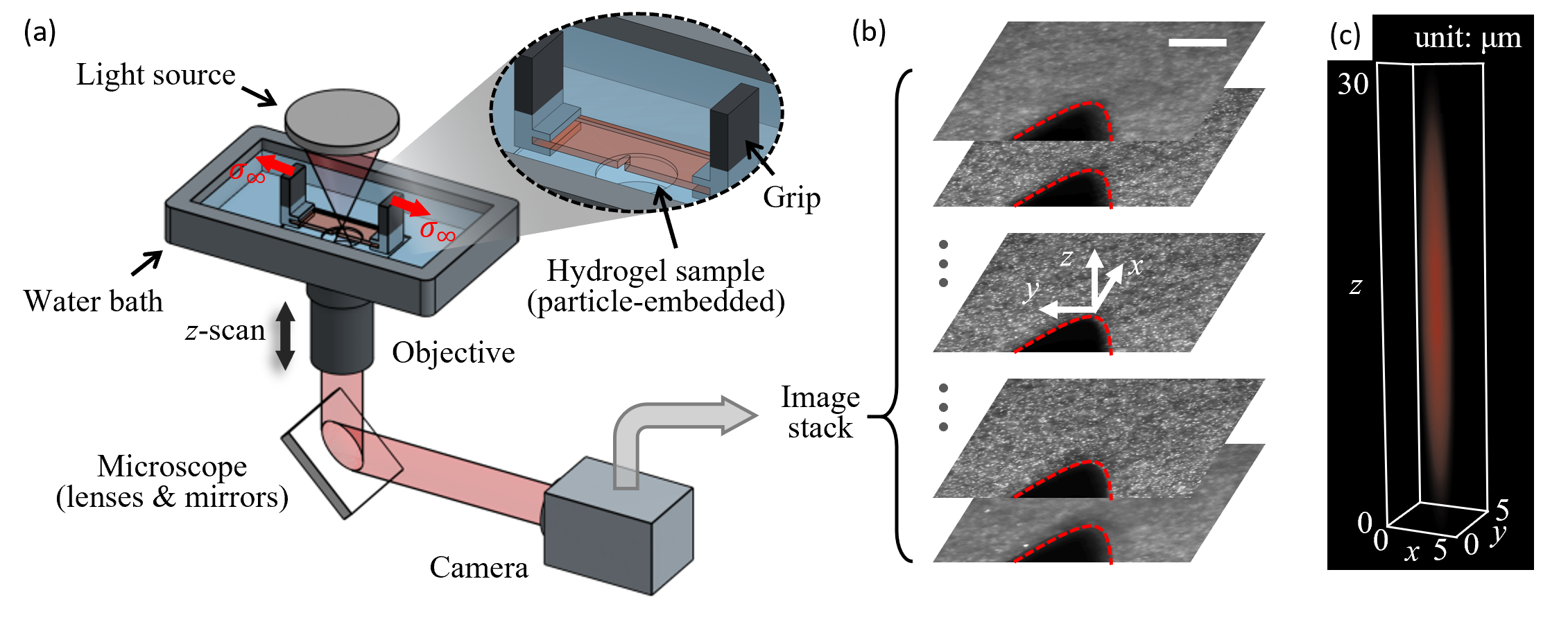}
\caption{(a) Experimental setup. A hydrogel sample is stretched by two symmetrically actuated grips. The hydrogel sample is $\SI{3}{\centi m}$ long, $\SI{1}{\centi m}$ wide, and $\SI{450}{\micro m}$ thick. The hydrogel was embedded with with passive particles, and a pre-cut was made before the uniaxial loading. During the experiment, light scattered from the particles is collected by a water-immersion objective (mounted on a microscope, not depicted) and transmitted to a high-resolution camera. By synchronizing the $z$-positioning of the objective and the camera image acquisition, volumetric image stacks are obtained. Inset: a magnified view of the hydrogel sample under stretch.
(b) An example of an recorded image stack. The sample was stretched along $y$-axis, and the crack propagates along $x$-axis. The crack surface is annotated by red dashed curve. The scale bar is $\SI{200}{\micro m}$ in $xy$-plane, and the $z$-spacing between slices is $\SI{5}{\micro m}$.
(c) 3D view of the light scattered by a single particle, as a representation of the point spread function of the imaging system. The light intensity is encoded by the color.}
\label{fig:setup}
\end{figure*}

Experiments are carried out using the experimental setup shown in Fig.~\ref{fig:setup}(a). The hydrogel sample is mounted on the grips of a home-built testing apparatus, and an edge crack (c.a. $\SI{2}{mm}$) is inserted on one side of the sample. The hydrogel sample is submerged in water throughout the experiment, and illuminated from a large angle using a light source to obtain a dark-field image. The light scattered from the particles is collected by a water-immersion objective ($10\times$, mounted on a Nikon TI eclipse microscope, not depicted) and imaged onto the sensor of a high-resolution camera (Hamamatsu C13440, resolution: $2048\times2048$ pixels, bit depth: 16 bit). By synchronizing the $z$-positioning of the objective and the image acquisition, volumetric image stacks are obtained. An example of the image stacks containing the polystyrene micro particles is shown in Fig.~\ref{fig:setup}(b), and the appearance of a single particle in the image stack is extracted and visualized in 3D in Fig.~\ref{fig:setup}(c). Because the lighting configuration collects scattering light, the particles appear larger than their actual size, with the apparent extent elongated along the $z$-axis.

In each experiment, a reference image stack capturing the center of the sample is recorded before any loading is applied. Then, by actuating a servo motor (not depicted), the grips move outwards symmetrically, and therefore, exert displacement-controlled remote tensile loading to sample. After each loading step, an image stack is recorded. During the loading, the pre-cut crack opens and starts to propagate slowly, eventually transiting to a smooth crack. At this stage, the displacement of the grips is held constant, and image stacks are recorded for further analysis as the crack propagates across the field-of-view.

\begin{figure*}[!htb]
\centering
\includegraphics[width=0.95\textwidth]{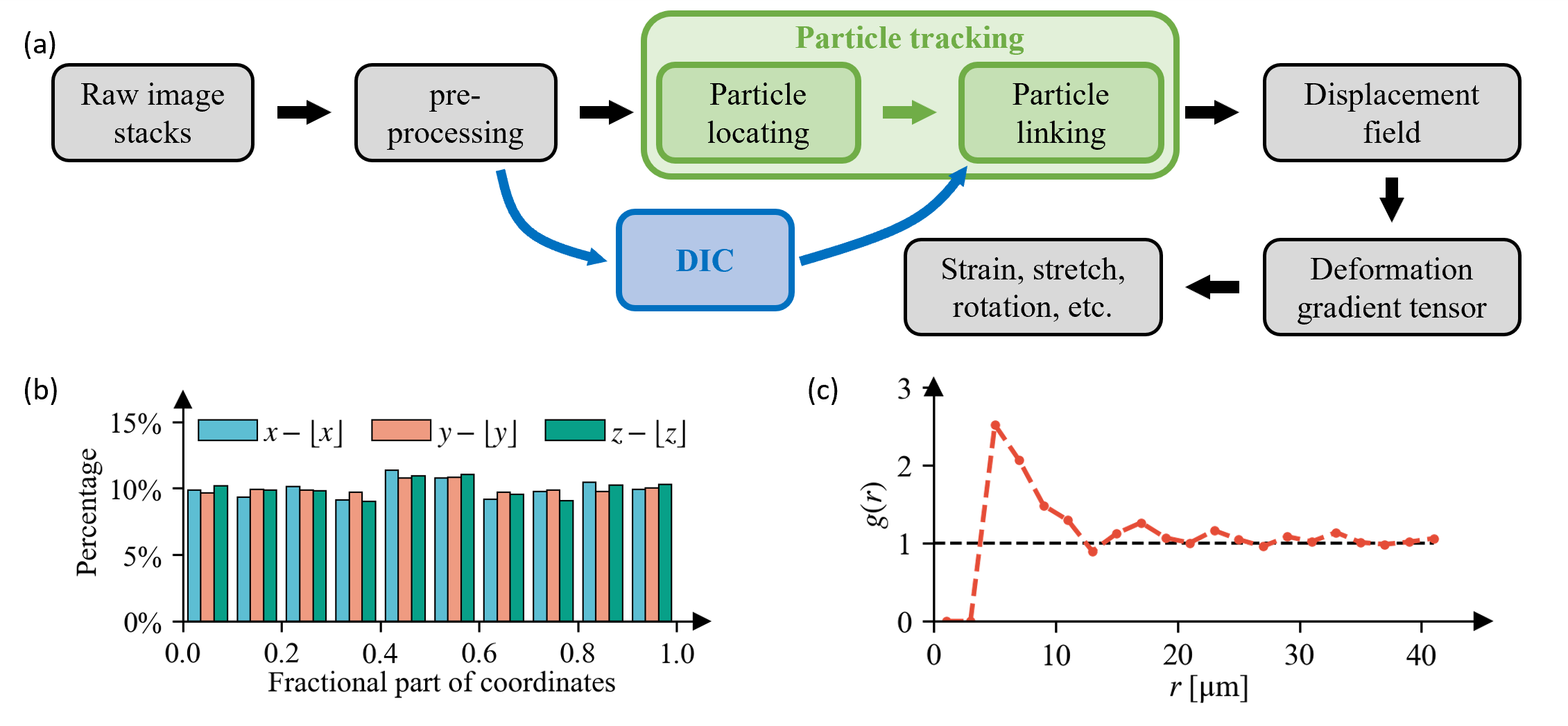}
\caption{(a) Image processing procedure. The raw image stacks are first pre-processed with a bandpass filter to eliminate most image noise, and then treated with 3D particle tracking algorithm to obtain the 3D displacement fields, which essentially consists of two steps - particle locating and particle linking. To elevate the reliability in linking particles, an intermediate DIC step is incorporated. Once the 3D displacement data is obtained, deformation gradient tensor is estimated at each particle location by a least-squares approach~\cite{Tyler_F_Estimator}, and further the kinematic data, such as stretch, rotation, and volumetric change.
(b) The histogram of the fractional part of particle coordinates in the reference frame, determined in the particle locating process. Given the random distribution of particles in the sample, the uniformity of the distribution validates the subpixel accuracy of the particle locations in all axes.
(c) The radial distribution function $g(r)$ of particles located in the reference frame. $g(r)$ peaks at approximately $\SI{8}{\micro m}$, and asymptotes to 1 after $\SI{15}{\micro m}$.
}
\label{fig:procedure}
\end{figure*}

The recorded image stacks are processed according to the workflow shown in Fig.~\ref{fig:procedure}(a) to obtain the 3D kinematic fields near the tip of the propagating crack. The raw image stacks are first pre-processed by a bandpass filter to suppress the image noise, and then analyzed with the open-source particle tracking algorithm, TrackPy~\cite{TrackPy}. In particle tracking, two essential steps are performed. First, particles are  located in each individual frame, and second, particles are linked between consecutive frames to identify their trajectories. During the locating phase, particles are identified in 3D in the image stacks, and their coordinates are extracted with subpixel resolution as the uniform distribution of the fractional part of particles’ coordinates plotted in Fig.~\ref{fig:procedure}(b) confirms. The distribution of the particles in the sample is statistically analyzed by the radial distribution function shown in Fig.~\ref{fig:procedure}(c), which implies that the average inter-particle distance is approximately $\SI{5}{\micro m}$ and the particles are uniformly distributed over a domain larger than $\SI{10}{\micro m}$. During the linking phase, an intermediate digital image correlation (DIC) step, using the open-source DIC software Ncorr~\cite{Ncorr}, is incorporated to enhance the linking reliability and accuracy. DIC calculation is sparsely carried out on selected slices in consecutive image stacks. Despite its limitations in handling large deformation and large rotation near the crack tip, DIC provides reliable predictions for particle linking in consecutive frames, where the deformation and rotation is significantly smaller. DIC ensures robust tracking even for particles that displace by a distance larger than the average inter-particle spacing between consecutive frames.

After successfully tracking the particles to form trajectories that can be traced back to the reference frame, full-field 3D displacements from the reference frame are directly calculated. The deformation gradient tensor $\mathbf{F}$ at each particle position is then estimated by a local least squares routine~\cite{Tyler_F_Estimator}, using the displacement vectors of nearest-neighbor particles. By performing the polar decomposition of $\mathbf{F}$, the near-crack-tip rotation tensor $\mathbf{R}$ and stretch tensor $\mathbf{U}$ are determined. Additionally, the local volumetric change of the hydrogel, i.e., the swelling ratio, is readily quantified by measuring the determinant of $\mathbf{F}$.

\section{Results}\label{sec:results}
Using the experimental method described in Sec.~\ref{sec:method}, we acquired image stacks of a propagating planar crack, nearly translationally invariant along $z$, and derived the particle trajectories near the crack tip. The image stacks have the size of $2048 \times 2048 \times 106$, with the resolution of $\SI{0.43}{\micro m/px}$ in $x$/$y$ and $\SI{5}{\micro m/px}$ in $z$. The crack was critically loaded and steadily propagated at a very slow speed of $\SI{0.02}{\micro m/s}$, as measured by the linear fit of the crack tip position as a function of time as plotted in Fig.~\ref{figS:crack_velocity}(a).

\subsection{Kinematic fields}\label{sec:kinematic}
\begin{figure*}[!htb]
\centering
\includegraphics[width=0.95\textwidth]{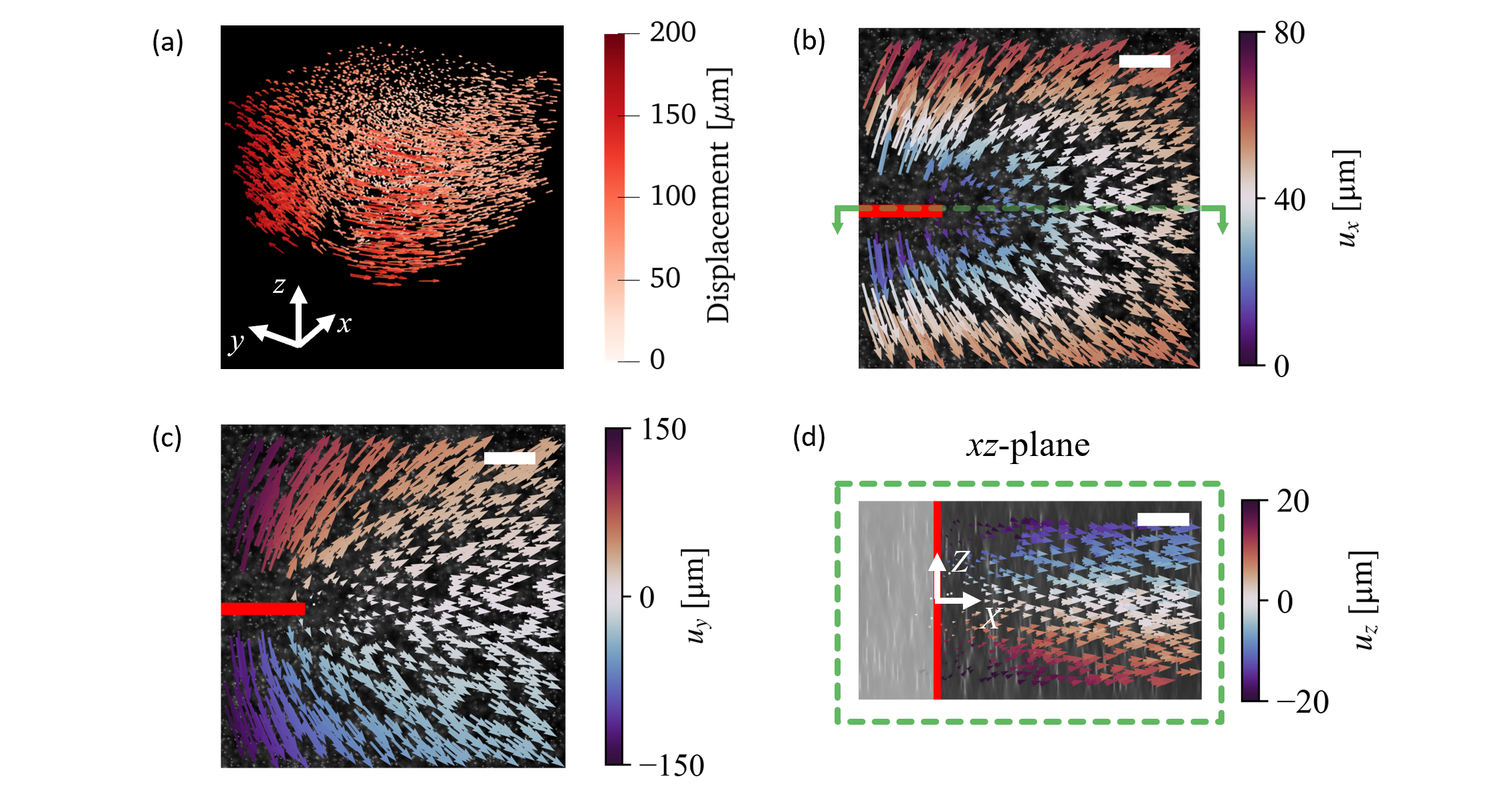}
\caption{(a) 3D displacement quiver plot for the particles tracked in the entire image stack (downsampled by a factor of 3 for better visualization). The arrows point from the particle locations in reference frame to their location in the current frame, and their length is scaled by 0.5. The arrows' color corresponds to the magnitude of displacement.
(b) and (c) 2D displacement quiver plots for the particles in the middle $xy$-plane (within $\pm \SI{10}{\micro m}$). The arrows are real scaled and color-coded by the magnitude of $u_x$ in (b) and $u_y$ in (c). The red solid line indicates the crack, and the green dashed line indicates the $xz$-plane that is shown in (d).
(d) 2D displacement quiver plot for the particles on the crack front $xz$-plane (within $\pm \SI{10}{\micro m}$). The arrows are real scaled and color-coded by the magnitude of $u_z$. The Poisson's effect leads to the displacement $u_z$ towards the sample's middle plane. The crack front is represented by the red solid line, and its faded left side represents where the crack opens after loading. The background image is resliced from the image stack and equally scaled in both axes.
The scale bar in (b)-(d) is $\SI{100}{\micro m}$.}
\label{fig:disp_fields}
\end{figure*}

% \subsection{Displacement fields}\label{sec:disp}
Displacement fields at each frame are readily calculated from particle trajectories for all three axes. A typical displacement field in our fracture experiment is visualized in 3D in Fig.~\ref{fig:disp_fields}(a). Representative particles are selected in the middle plane of the sample, and the displacement fields $u_x$ and $u_y$ for these particles are plotted in the reference frame in Fig.~\ref{fig:disp_fields}(b) and (c), respectively. As can be seen from the plots, particles are tracked very close to the crack tip. Both the direction and the magnitude of particle displacements are symmetric about the crack path, as consistent with the displacement predicted by LEFM for mode~I fracture. To investigate the thickness dependence of the displacement fields, we interrogate the particles in the $xz$-plane across the crack front; their displacement field is plotted in Fig.~\ref{fig:disp_fields}(d). According to the direction of the displacement vectors, the in-plane displacement $u_x$ is larger than the out-of-plane displacement $u_z$ for most of the particles. Nevertheless, a substantial $u_z$ due to material contraction along the $z$-axis is evident, especially near the top and bottom free surfaces at the crack tip.

\begin{figure*}[!htb]
\centering
\includegraphics[width=0.95\textwidth]{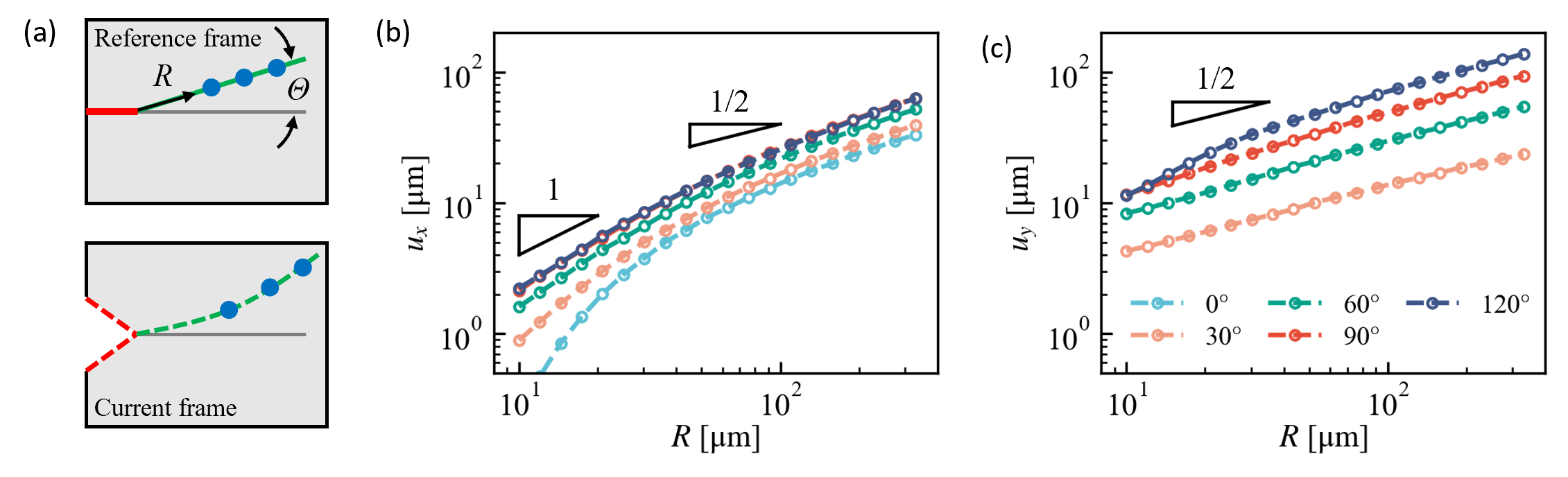}
\caption{(a) Data are interpolated at the points $(R, \Theta)$ in the middle $xy$-plane in reference frame (blue dots), where $R$ is the distance from the crack tip to the interpolation point and $\Theta$ is the angle of this line (represented in green) to the $x$-axis. Upon loading, the crack opens (red solid line to red dashed line), and the particles displace from their reference position (laid on green solid line) to their current position (laid on green dashed line).
(b) $u_x$ is interpolated in the reference frame at points on the lines with constant angles. The interpolation is done for total 53 frames and $u_x$ is averaged in time and shown. For a given $\Theta$, $u_x$ linearly increases with $R$ in the closest vicinity to the crack tip, and the increasing rate gradually reduces to $\sqrt{R}$ at slightly further distance. For a given $R$, $u_x$ increases with $\Theta$ until $90^{\circ}$. 
(c) $u_y$ is interpolated and shown in the same way. $u_y$ increases at a rate of $\sqrt{R}$ within the entire interrogated region ($\SI{330}{\micro m}$). Note that at $\Theta=0$, $u_y$ approaches zero, and therefore, is not visible in the log-log plot.}
\label{fig:disp_lines}
\end{figure*}

To analyze the displacement fields quantitatively, we evaluated the displacement components for material points along radial traces emanating from the crack tip in the material frame of reference for 5 different angles, as shown schematically in Fig.~\ref{fig:disp_lines}(a). $u_x$ and $u_y$ are interpolated at these points for a total of 53 frames, and the averaged values are represented in Fig.~\ref{fig:disp_lines}(b) and (c), respectively. The standard deviations are given in Fig.~\ref{figS:std_disp}. From the log-log plots, it is seen that for all evaluated angles, both $u_x$ and $u_y$ have the same trend with respect to $R$. Specifically, $u_x$ increases linearly with $R$ in the immediate vicinity of the crack tip ($R<\SI{50}{\micro m}$) but follows the $\sqrt{R}$ rate beyond $R \approx \SI{50}{\micro m}$; in contrast, $u_y$ follows the $\sqrt{R}$ trend across the entire investigated region.

\begin{figure*}[!htb]
\centering
\includegraphics[width=0.95\textwidth]{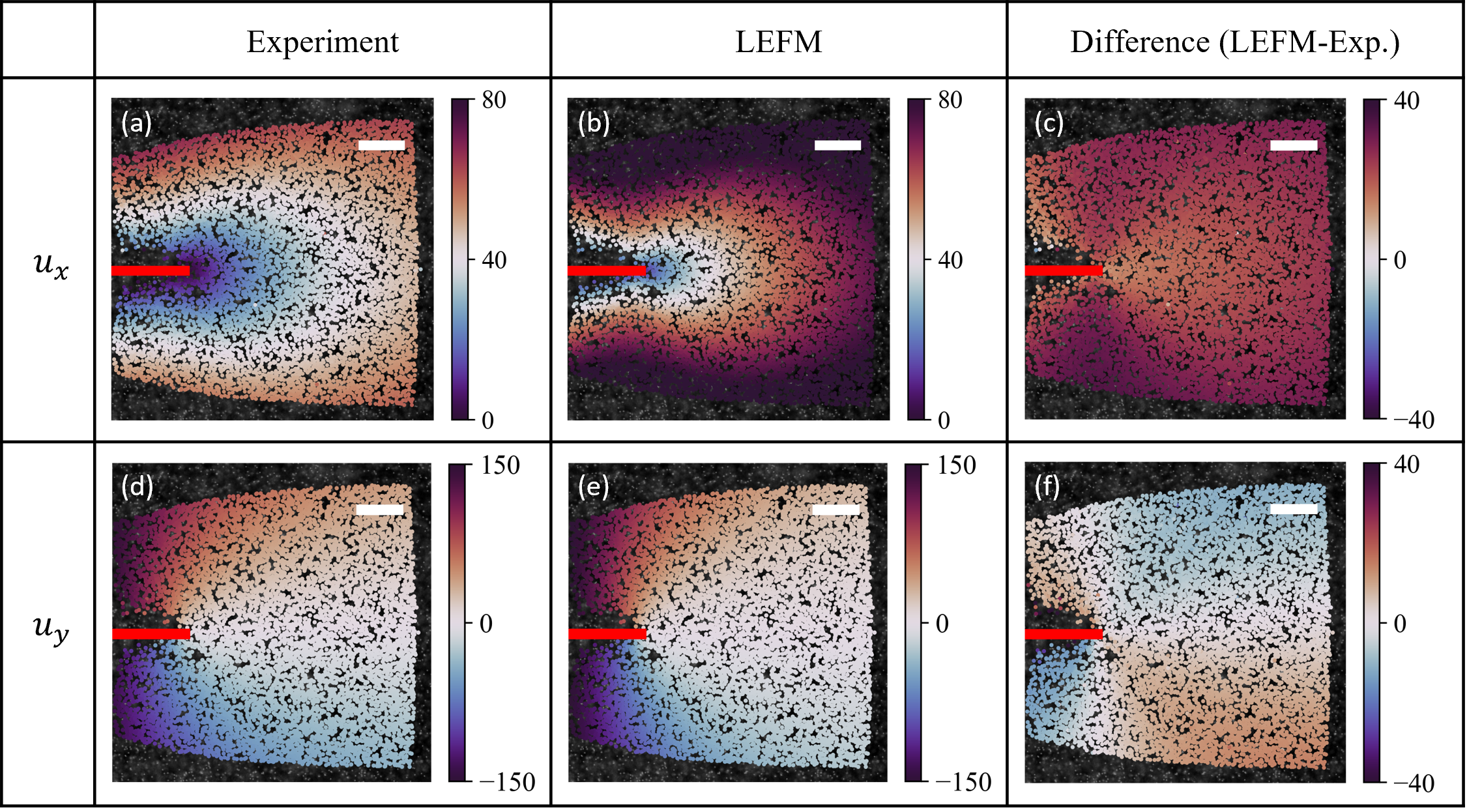}
\caption{Comparison of the measured in-plane displacement fields and the LEFM prediction~\cite{anderson2005fracture} at particles in the middle $xy$-planes (within $\pm \SI{100}{\micro m}$), (a)-(c) $u_x$ and (d)-(f) $u_y$. The difference shown in (c) and (f) is defined as the deviation of the LEFM prediction from the experimental measurement. The red solid line indicates the crack in the reference frame, and the white scale bar is $\SI{100}{\micro m}$. An evident over-prediction of $u_x$ is seen from the comparison.
% For $u_x$, LEFM over-predicts the displacement values across the entire measured region, and exhibits a more apparent displacement gradient. While for $u_y$, LEFM slightly under-predicts in the region ahead of the crack tip and slightly-over-predicts in the crack wakes. Despite the the larger values of $u_y$ comparing to $u_x$, the LEFM prediction of $u_y$ deviates less from the measurement.
}
\label{fig:disp_lefm}
\end{figure*}

The measured in-plane displacement components, $u_x$ and $u_y$, are compared with the LEFM prediction for Mode~I crack~\cite{anderson2005fracture}, as shown in Fig.~\ref{fig:disp_lefm}. The LEFM displacements are calculated with plane stress condition, a shear modulus of $\SI{35}{kPa}$~\cite{Livne2005, goldman2010}, and a Poisson's ratio of $0.48$ (representing the incompressibility of the hydrogel). The stress intensity factor is derived from the critical energy release rate $\SI{5}{J/m^2}$~\cite{Livne2005, goldman2010}. 

Generally from the comparison, it is seen that LEFM over-predicts the $u_x$ displacements across the entire measured field-of-view, but provides a relatively more accurate prediction for $u_y$, despite the magnitude of $u_y$ is larger than $u_x$ in the majority of the region. In the $u_x$ field, a constant offset approximately $\SI{25}{\micro m}$ is observed. Note that this offset is not due to the `\textit{T-stress}' during the experiment~\cite{bouchbinder2008weakly, livne2008breakdown}, as T-stress is a constant tensile stress which should introduce extra positive displacement in $u_x$ with a gradient. In the $u_y$ field, LEFM slightly under-predicts in the region ahead of the crack tip and slightly over-predicts in the wakes of the crack.

% \subsection{Rotation}\label{sec:rotation}
With the measured 3D particle displacement fields, the deformation gradient tensor $\mathbf{F}$ for each particle is calculated using a least squares approach~\cite{Tyler_F_Estimator}. Using the polar decomposition $\mathbf{F}=\mathbf{RU}$, the rotation tensor $\mathbf{R}$ and stretch tensor $\mathbf{U}$ are calculated.

\begin{figure}[!htb]
\centering
\includegraphics[width=0.45\textwidth]{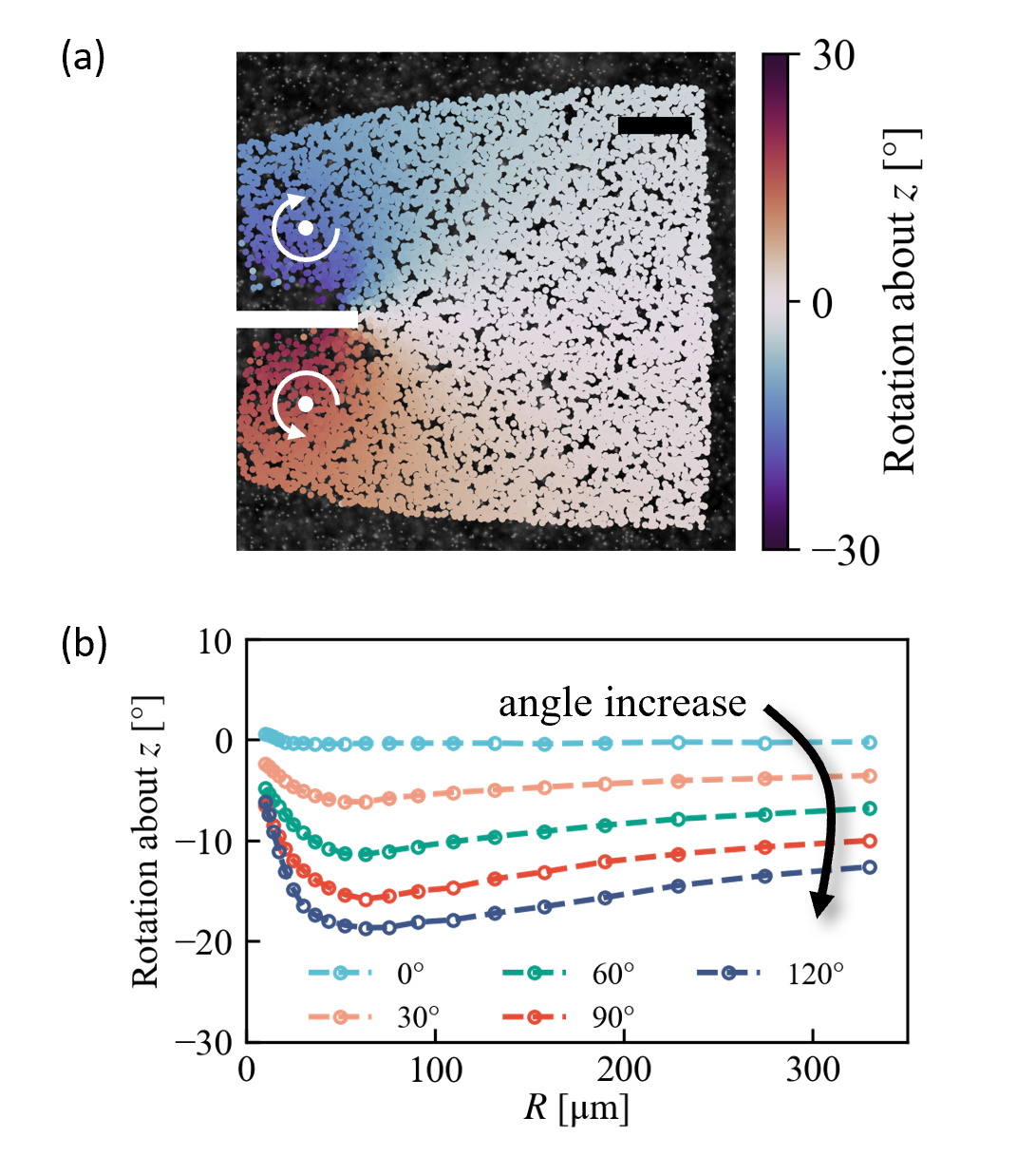}
\caption{(a) Rotation about $z$-axis, $R_z$, at particles in the middle $xy$-planes (within $\pm \SI{100}{\micro m}$). The rotation angle is calculated from the rotation tensor, which is obtained by the polar decomposition of the deformation gradient tensor ($\mathbf{F}$). The rotational effect about $z$ is minimal in the cone-shaped region ahead of the crack tip, but becomes prominent on both sides of the crack, especially after the crack just breaks the material.
The pair of circular arrows serve as the guide to the eye for the direction of the rotation. The scale bar is $\SI{100}{\micro m}$.
(b) Interpolation of $R_z$ at points same as Fig.~\ref{fig:disp_lines}. On the crack path ($\Theta=0$), the rotation about $z$ is approximately zero. For each interrogated line, the rotation peaks at $R$ around $50$ to $\SI{100}{\micro m}$.}
\label{fig:rotation_z}
\end{figure}

Rotation about the $X$, $Y$, and $Z$ axes is determined  from the rotation tensor $\mathbf{R}$~\cite{slabaugh1999computing}. The rotation about $Z$-axis is found to be the dominant component, and is plotted in Fig.~\ref{fig:rotation_z}(a). This rotation is anti-symmetric about the crack path, and becomes pronounced in the wake of the crack. The maximum rotation exceeds $30 \deg$. Unlike the rotation about the $Z$-axis, the rotation about the $X$- and $Y$-axes is small, as shown in the supplementary Fig.~\ref{figS:out-of-plane-rotation}.

We evaluate the rotation about the $Z$-axis at the same set of points illustrated in Fig.~\ref{fig:disp_lines}, with the average values computed over 53 frames depicted in Fig.~\ref{fig:rotation_z}(b) and the standard deviations in Fig.~\ref{figS:std_rotZ}. Along the crack propagating path ($\Theta=0$), the rotation is approximately zero. For all other angles evaluated, the magnitude of the rotation first increases with distance $R$ from the crack tip, and then gradually decrease to a finite value within the field-of-view. At a given distance $R$, the magnitude of the rotation monotonically increases with the angle.

% \subsection{Stretch fields}\label{sec:stretch}

\begin{figure*}[!htb]
\centering
\includegraphics[width=0.95\textwidth]{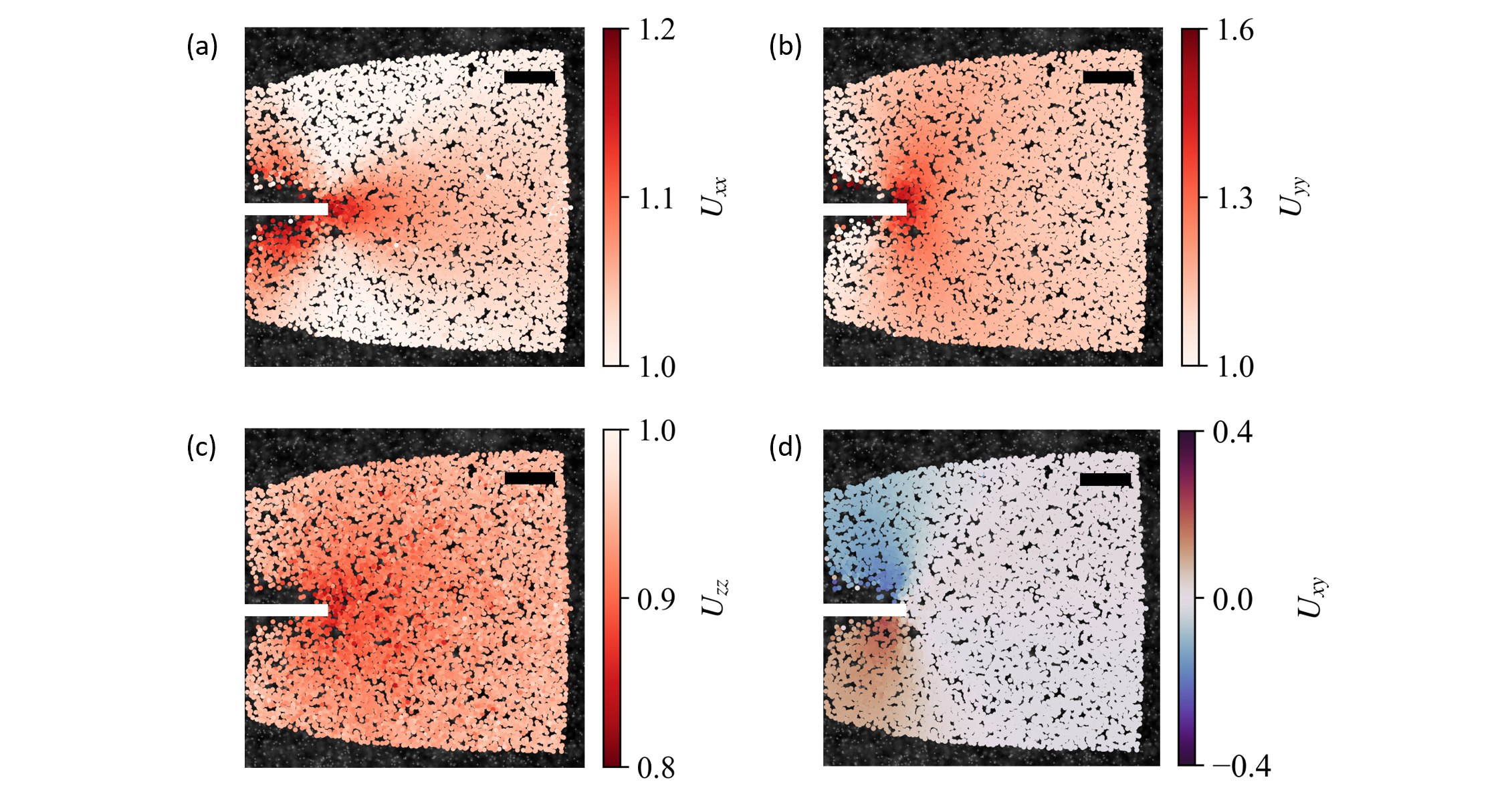}
\caption{The components of the right stretch tensor (a) $U_{xx}$,  (b) $U_{yy}$, (c) $U_{zz}$, and (d) $U_{xy}$, are plotted in the reference frame. The scale bar is $\SI{100}{\micro m}$. 
$U_{xx}$ is significant in a cone-shaped region ahead of the crack tip and near the crack surface. Stretch in the loading direction, $U_{yy}$, is dominant in the uncracked material, and exceeds 1.6 at the crack tip; near the crack surface, $U_{yy}$ gradually reduces to 1. Due to the Poisson's effect, contraction in $z$-axis is evident in the entire field of view, indicated by the $U_{zz}$ values less than 1. In-plane shear stretch $U_{xy}$ is minor in a broad region ahead of the crack tip, but develops rapidly post-crack, particularly near the crack surface.}
\label{fig:stretch_fields}
\end{figure*}

The stretch tensor $\mathbf{U}$ is also obtained from the polar decomposition, and the spatial distribution of the components $U_{xx}$, $U_{yy}$, $U_{zz}$, and $U_{xy}$ is illustrated in Fig.~\ref{fig:stretch_fields}. Despite the mode~I loading symmetry applied along the $Y$-axis, a substantial tensile stretch $U_{xx}$ is observed, particularly near the crack surfaces and in a cone-shaped region ahead of the crack tip. $U_{yy}$ exhibits stretch values higher than other components, and it intensifies when approaching the crack tip from any directions. 
$U_{zz}$ shows a more uniform distribution compared to the other components, with values typically less than one; these values suggest that the material near the crack tip generally experiences a compression across the thickness due to conservation of volume. The shear component is found to be minimal in the uncracked material, but becomes significant in the wake of the crack.

\begin{figure*}[!htb]
\centering
\includegraphics[width=0.95\textwidth]{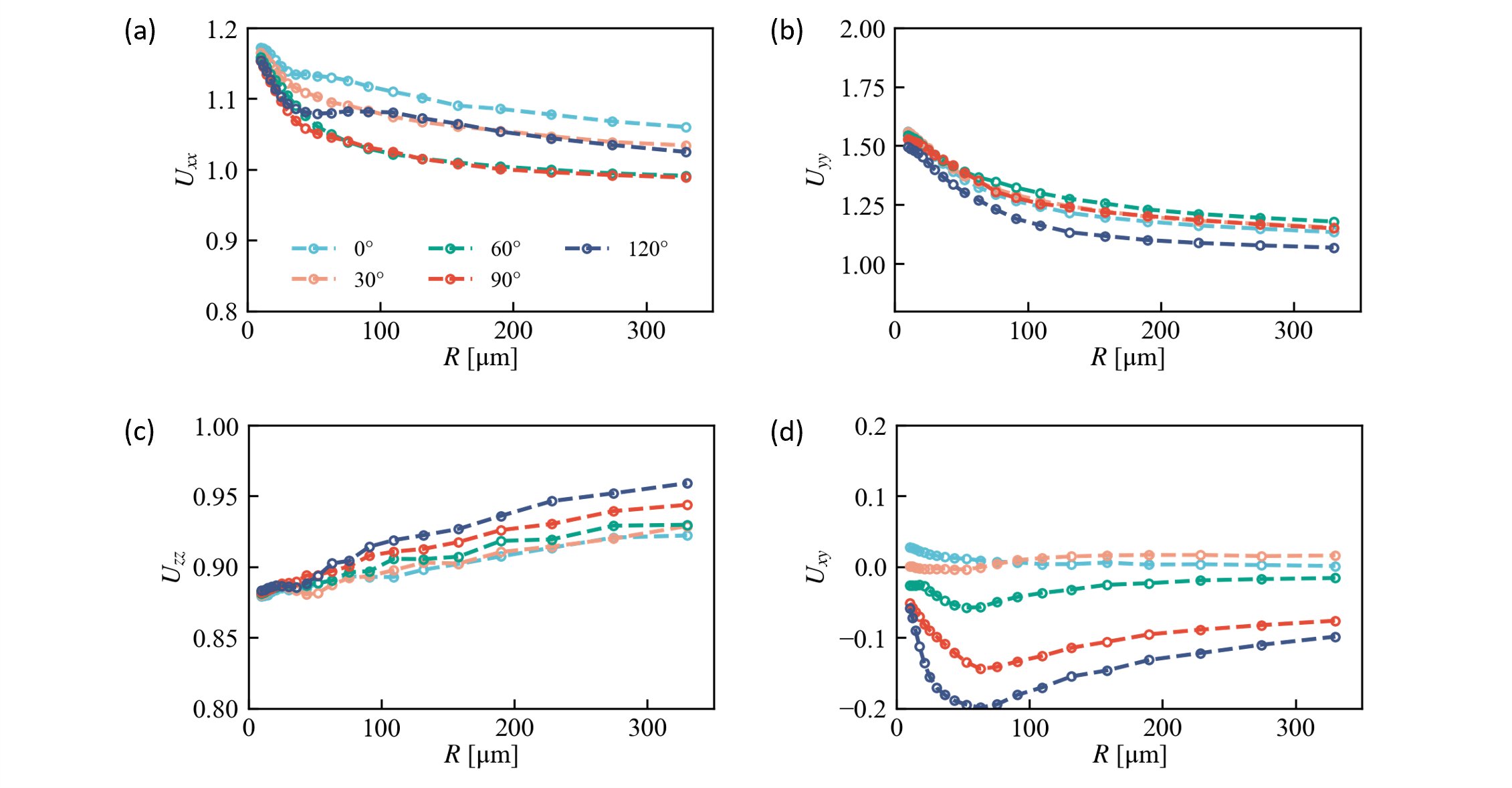}
\caption{Interpolated stretch values (a) $U_{xx}$, (b) $U_{yy}$, (c) $U_{zz}$, and (d) $U_{xy}$ in the reference frame, with the interpolation points and averaging frames consistent with those in Fig.~\ref{fig:disp_lines}.
In general, both $U_{xx}$ and $U_{yy}$ decrease as $R$ increases. However, due to the structure of $U_{xx}$ shown in Fig.~\ref{fig:stretch_fields}(a), the relation between $U_{xx}$ and $R$ is no longer strictly monotonic at large values of $\Theta$ near the crack tip; for a given $R$, $U_{xx}$ initially decreases with $\Theta$ until approximately $90^{\circ}$ and then increases.
The value of $U_{zz}$ becomes smaller as approaching the crack tip, indicating increasing contraction; $U_{zz}$ also exhibits more fluctuation, attributed to the lower image resolution in $z$ ($\SI{5}{\micro m/px}$) than that in $x$ and $y$ ($\SI{0.43}{\micro m/px}$).
The magnitude of $U_{xy}$ initially increases with $R$ until reaching approximately $\SI{50}{\micro m}$, after which it gradually decreases.
}
\label{fig:stretch_lines}
\end{figure*}

The stretch fields are quantitatively analyzed similar to the displacement fields along radial traces in the material frame of reference. The resulting average stretch values are plotted in Fig.~\ref{fig:stretch_lines}, with the standard deviations given in Fig.~\ref{figS:std_stretch}. For the in-plane stretch components, $U_{xx}$ and $U_{yy}$, the tensile stretch increases close to the crack tip, and the increasing slope becomes steeper and steeper, as expected due to the stress concentration near the crack tip. $U_{xx}$ shows very similar curves for the evaluation at $\Theta=60^{\circ}$ and $\Theta=90^{\circ}$, and the stretch values at the same radial distance $R$ are smaller than that at other angles, which is consistent with the structure of the $U_{xx}$ field shown in Fig.~\ref{fig:stretch_fields}. $U_{yy}$ has weak dependence on the angular position, and all curves converge on a finite far-field stretch value. The amplitude of the through-thickness compressive stretch, $U_{zz}$, consistently intensifies approaching the crack tip, seemingly in a linear relation with $R$ for all $\Theta$. The value of the shear component $U_{xy}$ is around zero for $\Theta\leq60^{\circ}$; for a larger $\Theta$, $U_{xy}$ first steeply increases with $R$ and maximize around $R=\SI{70}{\micro m}$, and then gradually decreases to a finite value. This is consistent with the large displacements along the crack tip opening.

Note that unlike the displacement fields, no angle-independent functional form is found to be potentially informative for the stretches; instead, the spatial structure in the stretch reflects the complication of the near-crack-tip deformation fields, as evidenced by the local multi-axial loading condition with finite values of $U_{xx}$, $U_{yy}$ and $U_{zz}$.

\begin{figure*}[!htb]
\centering
\includegraphics[width=0.95\textwidth]{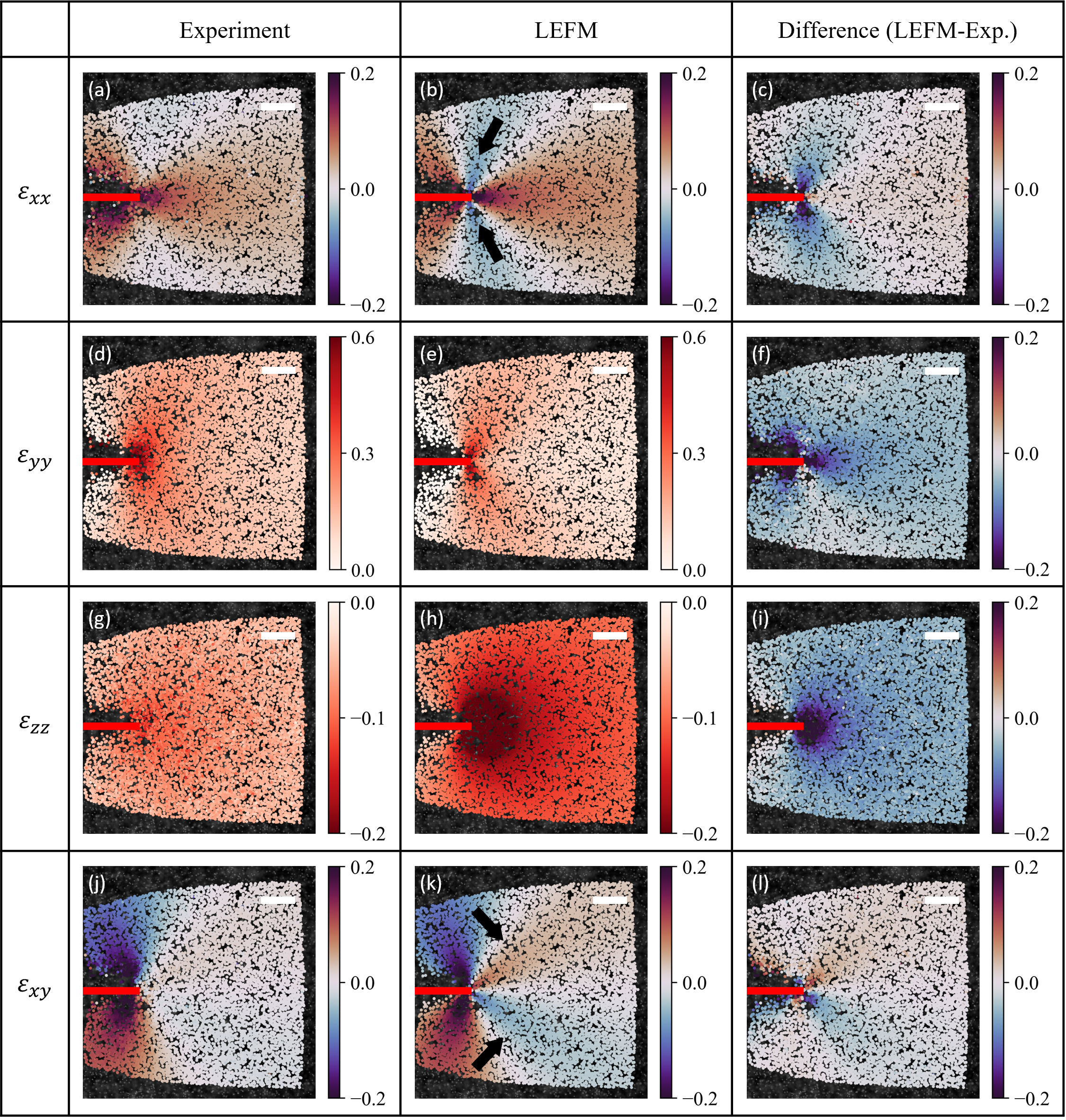}
\caption{Comparison of the measured strain fields and the LEFM prediction~\cite{anderson2005fracture} at particles in the middle $xy$-planes (within $\pm \SI{100}{\micro m}$), (a)-(c) $\varepsilon_{xx}$, (d)-(f) $\varepsilon_{yy}$, (g)-(i) $\varepsilon_{zz}$, and (j)-(l) $\varepsilon_{xy}$. The difference shown in the last column plots is defined as the deviation of the LEFM prediction from the experimental measurement. The red solid line indicates the crack in the reference frame, and the white scale bar is $\SI{100}{\micro m}$. The black arrows in (b) and (k) indicate the structures that are predicted by LEFM but do not show in the experiment. The LEFM prediction deviates from the measurement when approaching the crack tip, specifically in these regions for $\varepsilon_{xx}$ and $\varepsilon_{xy}$, along the crack path for $\varepsilon_{yy}$, and in the majority of near-crack-tip region for $\varepsilon_{zz}$.
}
\label{fig:strain_lefm}
\end{figure*}

We calculate the Green-Lagrange stain tensor from the stretch tensor by $\mathbf{E}=\frac{1}{2}\left[ \mathbf{U}^2 - \mathbf{I} \right]$, and compare the strain components ($\varepsilon_{xx}$, $\varepsilon_{yy}$, $\varepsilon_{zz}$, and $\varepsilon_{xy}$) to the LEFM predictions, as shown in Fig.~\ref{fig:strain_lefm}. The LEFM strain fields are calculated with plane stress condition and the same material properties used in the Fig.~\ref{fig:disp_lefm}. 

From the comparison, LEFM predicts the $\varepsilon_{xx}$ fields in good agreement with the experimental measurement, except the two cone-shaped structures indicated by the black arrows in the LEFM $\varepsilon_{xx}$ field shown in Fig.~\ref{fig:strain_lefm}(b). $\varepsilon_{yy}$ concentrates at the crack tip in both experiment and LEFM with similar values, but the shape is slightly different; this strain concentration appears to be more isotropic in the experiment than in the LEFM. In LEFM, the value of $\varepsilon_{yy}$ ahead of the crack tip is apparently smaller than at approximately $\pm 60^{\circ}$ to the crack path, leading to a substantial deviation in the region ahead of the crack tip. Approaching to the crack tip, this deviation becomes large. The through-thickness strain $\varepsilon_{zz}$ is predicted by LEFM significantly higher than the measurement across the entire field-of-view, especially at the crack tip, This is because LEFM calculates $\varepsilon_{zz}$ only as the contraction induced by the in-plane stress components, but in fact the relaxation due to solvent migration is not negligible as will be discussed in Sec.~\ref{sec:swelling}, particularly under the substantial $\varepsilon_{xx}$ and $\varepsilon_{yy}$ strains. The in-plane shear strain $\varepsilon_{xy}$ is overall well predicted by LEFM, except in the structures that are indicated by the black arrows in Fig.~\ref{fig:strain_lefm}(k).

\subsection{Near-crack-tip solvent transport}\label{sec:swelling}
We evaluate the near-crack-tip solvent transport by analyzing the local volume change within the material, which is readily obtained by computing the determinant of $\mathbf{F}$. This is a direct measurement of the swelling at the crack tip, straightforward but only feasible with fully resolved 3D kinematic fields; planar analyses yield valuable insights~\cite{Rong2019Fields, li2023pz}, but do not resolve the out-of-plane deformation.

\begin{figure*}[!htb]
\centering
\includegraphics[width=0.95\textwidth]{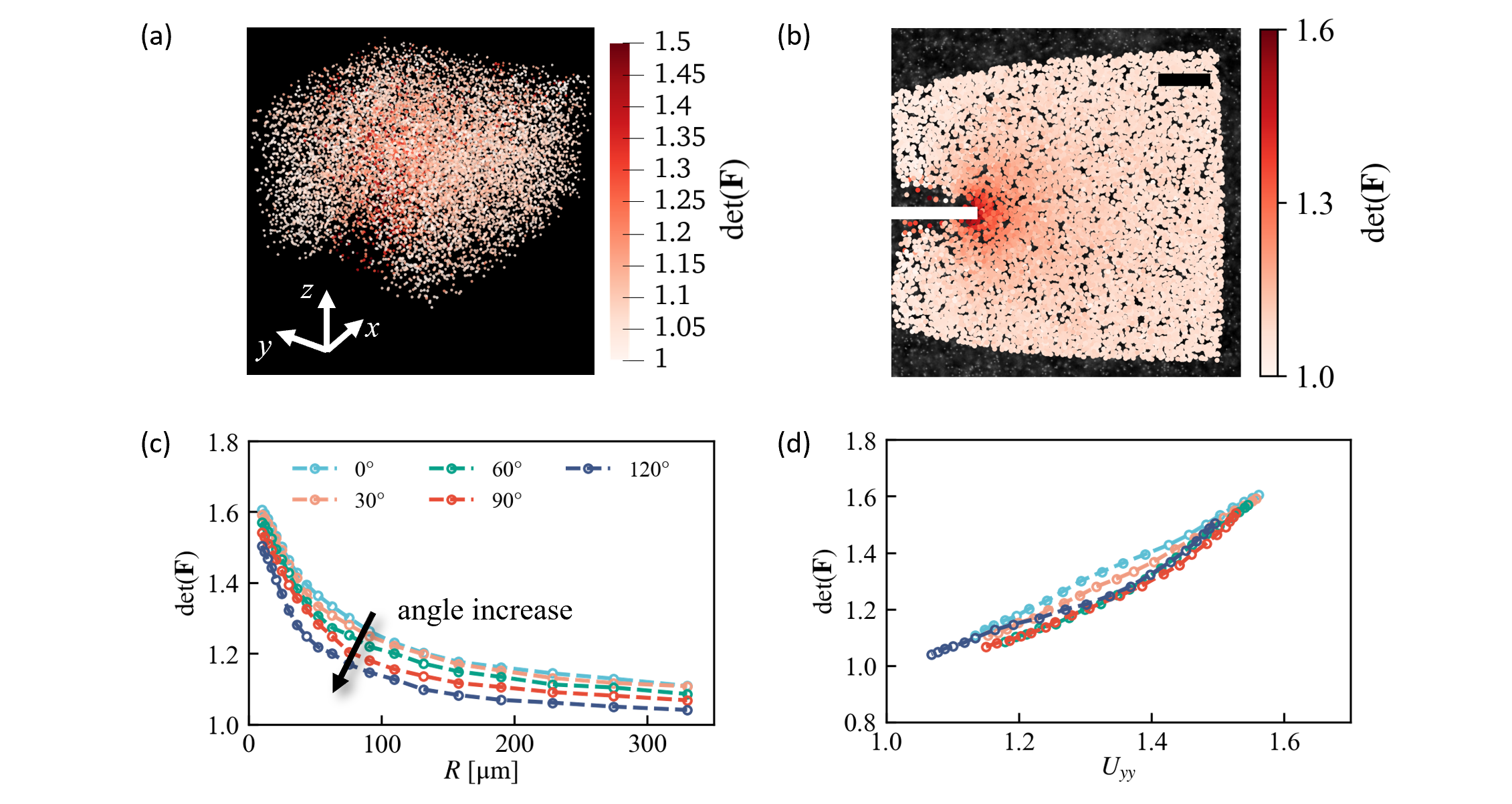}
\caption{(a) 3D visualization of the volumetric change near the tip of a fully-equilibrium propagating crack, depicted by the $\mathrm{det}(\mathbf{F})$.
(b) Distribution of $\mathrm{det}(\mathbf{F})$ in the middle $xy$-plane (within $\pm \SI{100}{\micro m}$). 
(c) Interpolation of $\mathrm{det}(\mathbf{F})$  (same procedure in Fig.~\ref{fig:disp_lines}). The volumetric change intensifies as approaching the crack tip, as well as the rate of this change. At points with a specific radial distance $R$, $\mathrm{det}(\mathbf{F})$ decreases when the angle increases.
(d) $\mathrm{det}(\mathbf{F})$ with respect to stretch $U_{yy}$ at the interpolation points, revealing a positive correlation between them.
}
\label{fig:swelling}
\end{figure*}

We calculate $\mathrm{det}(\mathbf{F})$ for all particles, and the distribution of the resulting values is visualized in 3D in Fig.~\ref{fig:swelling}(a). $\mathrm{det}(\mathbf{F})$ is observed to be generally greater than one over the entire volume, and a pronounced non-uniformity can be seen near the crack tip, which can be seen with greater clarity in the 2D projection shown in Fig.~\ref{fig:swelling}(b). In the immediate vicinity of the crack tip, $\mathrm{det}(\mathbf{F})$ is large, and can exceed 1.5, corresponding to a volume increase greater than $50\%$. Considering the incompressibility of the solvent (water) and polymer chains, this volumetric change arises solely due to solvent migration.

We analyze $\mathrm{det}(\mathbf{F})$ at groups of points along radial traces emanating from the crack tip, similar to the analysis for the kinematic fields as can be seen in Fig.~\ref{fig:swelling}(c) the average values and in Fig.~\ref{figS:std_swelling} the standard deviations. Along all the $\Theta$ evaluated, $\mathrm{det}(\mathbf{F})$ sharply increases close to the crack tip, and the increasing rate reduces when $R$ increases. This changing rate also depends on the angular position of the evaluation point; for a given $R$, $\mathrm{det}(\mathbf{F})$ changes more drastically for a larger $\Theta$ than for a small $\Theta$, particularly when $R<\SI{100}{\micro m}$; this indicates a concentration of solvent migration in the region ahead of the propagating crack that is not uniformly distributed, but localized on the crack axis. These curves are very similar to the stretch curves $U_{yy}$ shown in Fig.~\ref{fig:stretch_lines}(b) in terms of shape. In fact, a strong correlation is evidenced between $\mathrm{det}(\mathbf{F})$ and the dominant stretch $U_{yy}$, as illustrated in Fig.~\ref{fig:swelling}(d). This observation suggests that the swelling at the crack tip may be induced by stretch.

\begin{figure}[!htb]
\centering
\includegraphics[width=0.45\textwidth]{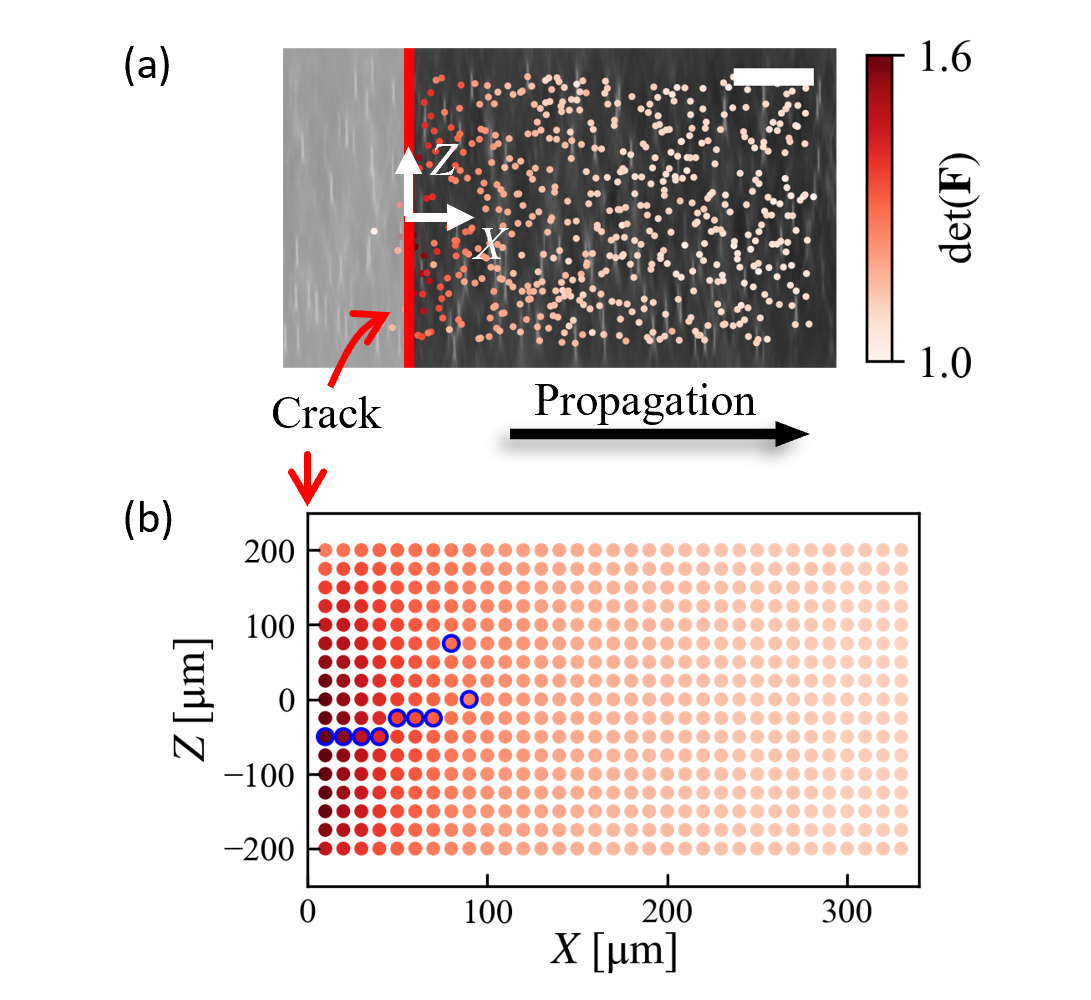}
\caption{(a) Distribution of $\mathrm{det}(\mathbf{F})$ on the crack front $xz$-plane (within $\pm \SI{10}{\micro m}$). The red solid line represents the crack front and the faded region represents to the crack opening region upon loading. The scale bar is $\SI{100}{\micro m}$. 
(b) Interpolation of $\mathrm{det}(\mathbf{F})$ is conducted on the crack front $XZ$-plane in the reference frame and over a total of 53 frames. The color-coded values of $\mathrm{det}(\mathbf{F})$ are displayed using the same color scale as shown in panel (a). Blue circles donate the locations of the maximum of $\mathrm{det}(\mathbf{F})$ at the same $X$ position. A noticeable variation in $\mathrm{det}(\mathbf{F})$ along the $z$ direction is observed near the crack front, peaking near the mid-plane of the sample. Beyond $\SI{100}{\micro m}$ from the crack front, $\mathrm{det}(\mathbf{F})$ becomes uniform in $z$ (with a standard deviation $\approx 0.006$).
% (c) The maximum $\mathrm{det}(\mathbf{F})$ at each $X$ position is depicted. The $\mathrm{det}(\mathbf{F})$ gradually decreases with increasing $X$, but still maintains a finite value even at $\SI{300}{\micro m}$ away from the crack tip.
}
\label{fig:swelling_evo}
\end{figure}

The distribution of $\mathrm{det}(\mathbf{F})$ is also studied in the crack plane ahead of the crack tip, as shown in Fig.~\ref{fig:swelling_evo}(a). The evolution of $\mathrm{det}(\mathbf{F})$ along $X$-axis remains consistent across all $Z$-position, where $\mathrm{det}(\mathbf{F})$ becomes larger as $R\rightarrow 0$. The distribution of $\mathrm{det}(\mathbf{F})$ at a given $X$ does not vary significantly along the $Z$-axis; however, after interpolating and averaging $\mathrm{det}(\mathbf{F})$ on a grid shown in Fig.~\ref{fig:swelling_evo}(b), a slight but evident $\mathrm{det}(\mathbf{F})$ difference in $Z$ is observed, particularly near the crack front. At each $X$-position, $\mathrm{det}(\mathbf{F})$ is found to maximize near the middle of the sample, as indicated by blue circles, and decreases to the free surfaces. This difference is presumably due to the $Z$ variation of the stress tri-axiality, which varies from nearly plane stress near the free surfaces to nearly plane strain in the middle region.
% The maxima are plotted in Fig.~\ref{fig:swelling_evo}(c) as a function of $X$. 

\section{Discussion}
In this work, we measure the kinematic fields near the tip of a slowly propagating planar crack and highlight the substantial swelling at the crack tip. These characterizations are carried out based on 3D particle trajectories measured from volumetric image stacks and the resulting particle tracking. The particle trajectories and derivative kinematic quantities are computed purely geometrically, without making any assumptions about the material. Therefore, this experimental method can be readily adopted for similar characterization in other soft materials~\cite{Dickey1, Long, Esther, Jin_elastomer, Zehnder_PDMS}. This analysis may be applied to questions in fundamental fracture mechanics, including fully 3D crack perturbations~\cite{hodgdon_sethna, rice94, movchan_gao_willis, lebihain}, 3D crack path selection criteria~\cite{erdogan_sih, goldstein, slepyan, amestoy_leblond, rozen2020fast}, the kinematics of crack surface patterns~\cite{wei2024complexity, wang2024size, wang2022hidden, tanaka1998discontinuous, baumberger2008magic, kolvin2018topological, ravi1984experimental, fineberg1991instability, sharon1996microbranching, Livne2005, goldman2010, Livne2010Science}, and crack stability under mixed-mode loading conditions~\cite{suresh, xu, lazarus, pons_karma, lin, pham1, pham2}.

The crack generating the fields analyzed in this manuscript is propagating extremely slowly - though the entire field of view is of the order of a mm or so, this experiment took$\approx$ 3 hours (see Fig.~\ref{figS:crack_velocity}(a)); this timescale is comparable to the poroelastic timescale, $B^2/D \approx  \SI{1600}{s}$, where $B = \SI{450}{\micro m}$ is the sample thickness and $D \approx \SI{e-10}{m^2/s}$ is the effective diffusivity of water molecule in a polyacrylamide hydrogel~\cite{kalcioglu2012macro}. This slow crack propagation results in significant and diverging swelling at the crack tip, where the swelling is highly correlated with the stretch along the loading direction. Indeed, the multi-component stretch state at the crack tip can lead to solvent migration, as evidenced in that uniaxial and biaxial tensile tests of hydrogels~\cite{takigawa1993uniaxial, galli, hu2010using, Urayama2015Biaxial}; on the other hand, the swelling, as a time-dependent process, can modify the local kinematic fields, with consequences for the local material properties and the stress fields~\cite{ hu2010using, kalcioglu2012indentation}. While the incompressible neo-Hookean material model has been extensively used in dynamic fracture experiments of hydrogels, where it is highly accurate due to the short time duration of the sample loading during crack propagation~\cite{Livne2005, Livne2010Science, rozen2020fast, goldman2010, wang2023tensile}, care should be taken in selecting material models for slow cracks.

In order to determine whether the observations here are time-dependent, we carried out a second experiment for a faster, albeit still very slow, crack ($v\approx \SI{0.63}{\micro m/s}$). The duration of this experiment is shorter than the poroelastic timescale, and therefore, the solvent does not have sufficient time to equilibrate. Direct comparison of the displacement fields, stretch fields, and swelling is provided in supplementary figures Fig.~\ref{figS:undrained_displacement}-\ref{figS:undrained_detF}. A similar displacement trend is found for both the equilibrated and non-equilibrated cracks. The in-plane stretch components $U_{xx}$ and $U_{yy}$ are similar for both cracks, but a disparity in $U_{zz}$ is observed. Near the crack tip, $U_{zz}$ for the equilibrated crack is larger than that for the non-equilabrated crack, indicating that the material at the crack tip relaxes through the thickness direction during the transition from the non-equilabrated state to the equilabrated state. This poroelastic relaxation contributes to the dilation of the material. The difference in $\mathrm{det}(\mathbf{F})$ for the equilabrated and non-equilabratd crack is also observed in Fig.~\ref{figS:undrained_detF}, consistent with poroelastic solvent migration at the crack tip.

The swelling near the crack tip is shown to be induced by the stretch, but it still remains uncertain through which component. As depicted in Fig.~\ref{fig:swelling}(d), $U_{yy}$ seems highly linearly correlated with $\mathrm{det}(\mathbf{F})$. But comparing the equilibrated and nonequilibrated cracks, the values of $U_{yy}$ are almost identical in the entire evaluated region, whereas a substantial difference is evident in $\mathrm{det}(\mathbf{F})$. Microscopically, the osmotic stress competes with the stress on the polymer chains; on the macroscopic scale, the poroelastic swelling may mutually affect the material's mechanical properties and stretch, making the coupled problem challenging.

Another interesting observation of the experiments is the extremely slow crack velocity that is nevertheless stable, despite the uniformity of the applied strain in the far-field. Highly cross-linked polyacrylamide hydrogels are canonically modelled as brittle material, analogy to the traditional brittle materials such as glass and ceramics; however, in these brittle materials, to the best of our knowledge, mode~I fracture has never been reported at this extremely slow speed. We suspect that in our experiments, the slow cracks may be stabilized by the poroelasticity~\cite{baumberger2020environmental}, possibly by means of viscous dissipation in the flow through the gel's polymer network. Indeed, we have never achieved this slow crack speed in air. Certainly, we can verify this hypothesis by changing the solvent surrounding the crack. But more in-depth measurement is required to identify the mechanism that stabilizes this slow crack growth.

While the 3D measurements provide rich kinematic information near the crack tip, the analysis of the stress is now limited by the constitutive model. As shown in the results, at the crack tip, not only is the deformation substantial at the crack tip, but the local loading condition is also multi-axial. On top of that, solvent migration adds poroelasticity to the material response. Additionally, distributed damages around the crack tip may also locally alter the material properties~\cite{li2023pz, deng2023nonlocal}. Despite these challenges, an accurate and physically-meaningful material model is indispensable to facilitate further stress analysis near the crack tip and to evaluate the local $J$-integral analysis~\cite{rice_1968}.

\section{Conclusion}
In this manuscript, we measured the fully-3D displacement fields near the tip of a slowly propagating crack in a hydrogel sample, using a microscopic 3D imaging technique and a DIC-assisted particle tracking algorithm. Using the high-resolution displacement fields, we estimated the deformation gradient tensor fields, and further obtained the near-crack-tip rotation fields, stretch fields, strain fields, and swelling fields. Our results uncover substantial material rotation, which can exceed $30^{\circ}$ in the wake of the crack. The local loading condition at the crack tip is complicated; the multiaxiality shown by the finite values of $U_{xx}$, $U_{yy}$, and $U_{zz}$; and the large stretch exceeding 1.6 in the loading direction. The displacement fields and strain fields are compared with the LEFM predictions for Mode~I crack; displacement component $u_x$ is found to be over-predicted by LEFM across the field-of-view; strain field are estimated over-structured in LEFM $\varepsilon_{xx}$ and $\varepsilon_{xy}$ fields, leading to inaccurate strain values in these regions; $\varepsilon_{zz}$ contraction is apparently over-estimated across the field-of-view, particularly at the crack tip. The fully resolved 3D displacement field enabled a quantitative measurement of the significant solvent migration at the crack tip. The solvent migration led to hydrogel swelling, which is highly correlated with the local stretch along the loading axis, and is dependent on the crack velocity. The extremely slow crack speed observed may result from poroelastic flow and viscous losses, but further study is required to test this conjecture. The experimental method we used can open a door for 3D characterization of near-crack-tip kinematic fields, particularly for complex cracks and for soft materials, where essential material structure approaches the micron-scale. Our experimental results are expected to provide insights for constitutive model development / selection for soft materials undergoing multi-axial loading and poroelastic swelling, and might prove useful for the validation of numerical calculations of hydrogel systems.

% \begin{acknowledgements}
% The research was funded by Swiss National Science Foundation Grant no. 200021\_197162.
% \end{acknowledgements}

\section*{Conflict of interest}

The authors declare that they have no conflict of interest.

% \bibliographystyle{spmpsci}      % mathematics and physical sciences
\bibliographystyle{spphys}       % APS-like style for physics

\bibliography{Ref}   % name your BibTeX data base

\end{document}